\documentclass[fleqn,usenatbib]{mnras}

\usepackage{newtxtext,newtxmath}

\usepackage[T1]{fontenc}

\DeclareRobustCommand{\VAN}[3]{#2}
\let\VANthebibliography\thebibliography
\def\thebibliography{\DeclareRobustCommand{\VAN}[3]{##3}\VANthebibliography}

\usepackage{ae,aecompl}
\usepackage{graphicx}	
\usepackage{amsmath}	
\usepackage{times}
\usepackage[usenames,dvipsnames]{xcolor}
\usepackage{soul}
\usepackage{comment}
\usepackage{tablefootnote}
\usepackage{threeparttable}

\title[Spectra of Brown Dwarf Candidates in the ONC]{JWST Spectra of Brown Dwarf Candidates in the Orion Nebula Cluster}

\author[K. L. Luhman]{
K. L. Luhman$^{1,2}$\thanks{E-mail: kll207@psu.edu}
\\
$^{1}$Department of Astronomy and Astrophysics, The Pennsylvania State University, University Park, PA 16802, USA\\
$^{2}$Center for Exoplanets and Habitable Worlds, The Pennsylvania State University, University Park, PA 16802, USA
}

\date{Accepted XXX. Received YYY; in original form ZZZ}

\pubyear{2025}

\begin{document}
\label{firstpage}
\pagerange{\pageref{firstpage}--\pageref{lastpage}}
\maketitle

\begin{abstract}
I present an analysis of archival spectra of 200 sources toward the
Orion Nebula Cluster (ONC) that were obtained with the
Near-Infrared Spectrograph (NIRSpec) on board the James Webb Space 
Telescope (JWST). I have used these data to assess cluster membership
and measure spectral types for the targets. Fifty-three sources are classified
as likely cluster members, 24 of which have spectral types that are
suggestive of brown dwarfs ($>$M6). 
Seven of the NIRSpec targets were previously identified as
``Jupiter-mass binary objects" (JuMBOs), all of which are background sources 
rather than brown dwarfs based on the NIRSpec data.
The spectral classifications of those objects are consistent with the
results of my recent study of the JWST photometry in the ONC, which found 
that only a few JuMBO components have the colors expected for brown dwarfs, 
none of which form pairs that have uniquely wide separations or low
masses relative to known binary brown dwarfs.
\end{abstract}

\begin{keywords}
brown dwarfs -- stars: formation 
\end{keywords}



\section{Introduction}
\label{sec:intro}

The Orion Nebula Cluster \citep[ONC,][]{mue08,ode08} has served
as a popular hunting ground for newborn brown dwarfs because of its
youth \citep[1--5 Myr,][]{jef11}, proximity \citep[$390\pm2$ pc,][]{mai22},
and richness ($\sim$2000 members). The compact spatial distribution 
of the ONC has made it amenable to deep imaging with ground-based 
telescopes \citep{mcc94,sim99,luc00,hil00,mue02,luc05,rob10,dar12,dra16,mei16}
and space-based facilities like the Hubble Space Telescope
\citep{luh00,and11,rob20,gen20}. Soon after the
James Webb Space Telescope \citep[JWST,][]{gar23} began science operations
in 2022, \citet{mcc23} used one of its cameras, NIRCam \citep{rie05,rie23},
to obtain images of a field in the center of the ONC that encompasses 
$\sim25$\% of the cluster members. Based on those images, \citet{mcc23} and 
\citet{pea23} (hereafter MP23 and PM23) reported the
detection of 540 candidate members with mass estimates below 13~$M_{\rm Jup}$, 
86 of which appeared in pairs and trios that were named ``Jupiter-mass binary 
objects" (JuMBOs).
Many of the JuMBOs had projected separations of $>$100 AU at the distance
of the ONC. The formation of binaries with such wide separations and low
masses would be challenging to explain. A number of subsequent studies have 
attempted to develop theories for the formation of the JuMBOs under
the assumption that they are bona fide substellar binaries
\citep{col24,dia24,hua24,laz24,wan24,wan25,yu24,par25}.

In \citet{luh24o2}, I performed a new search for brown dwarf candidates
in the JWST images of the ONC from MP23 and PM23.
I found that many of the JuMBOs are probably reddened background sources 
based on their colors while others have inadequate photometry for assessing 
their nature. The photometry suggested that only one promising substellar 
binary is present among the JuMBOs, and that its components do not approach
the mass of Jupiter.
MP23 and PM23 did not report coordinates for their 
single candidates, so it is not possible to assess their validity.

Spectroscopy is necessary for confirming brown dwarf candidates
like the JuMBOs in the ONC. JWST program 2770 (PI: M. McCaughrean)
has used the multiobject mode of the Near-Infrared Spectrograph
\citep[NIRSpec,][]{fer22,jak22} on JWST to obtain spectra of a few hundred
sources toward the ONC, including a subset of the JuMBOs.
In this Letter, I present an analysis of the data from that program that 
are publicly available.

\section{JWST/NIRSpec Observations}

JWST program 2770 observed a few hundred sources toward the ONC 
using the multiobject spectroscopy mode of NIRSpec on JWST
in February and September of 2024. The February data are
publicly available and are analyzed in this paper.
The observations employed the microshutter assembly (MSA) on NIRSpec,
which has four quadrants that each contain 365$\times$171 shutters.
Each shutter has a size $0\farcs2\times0\farcs46$.
The MSA spans an area of $3\farcm6\times3\farcm4$.
The instrument collects spectra with two $2048\times2048$ detector arrays that 
have pixel sizes of $0\farcs103\times0\farcs105$. NIRSpec was operated
with the PRISM disperser, which covers 0.6--5.3~\micron\ with a spectral
resolution of $\sim$40 to 300 from shorter to longer wavelengths.

The February observations were performed with five MSA configurations, each
in a separate visit. For a given configuration, data were collected at
three nod positions that moved targets between three adjacent shutters.
which were equivalent to $0\farcs2\times1\farcs5$ slitlets.
A single exposure was taken at each nod position, which utilized nine groups, 
eight integrations, and the NRSIRS2RAPID readout pattern.
The total exposure time for each configuration was 3545~s.

To reduce the NIRSpec data, I began by retrieving the {\tt uncal} files 
from the Mikulski Archive for Space Telescopes 
(MAST)\footnote{\url{https://doi.org/10.17909/v8xw-dc33}}.
Those files were processed with the JWST Science Calibration pipeline
version 1.17.1\footnote{\url{https://doi.org/10.5281/zenodo.14597407}}.
For each target, background subtraction for a given nod was performed
using the average of the other two nods, and the background subtracted
spectra from the three nods were combined. However, the background
subtraction in that initial reduction was not optimal for some targets
because of the spatial variation of nebular emission or the presence
of neighboring sources in the slitlets. Therefore, if the initial
spectrum of a target appeared to have spectral features expected of
a cluster member (Section~\ref{sec:class}) but the background subtraction
was poor, a given nod was background subtracted with each of the other
two nods separately, resulting in three options for background subtraction
for each of the three nods.  Those options were compared for a given nod
to identify the one that would be adopted. In some cases, only one or two of 
the nods had adequate background subtraction to be retained.
I did not attempt to identify the optimal reduction for the targets 
classified as background sources based on the initial reduction,
so emission or absorption at the wavelengths of nebular lines in those
reduced spectra should be treated with caution.  Even for targets that have 
refined background subtraction because they were classified as members,
subtraction of the background nebular emission lines is often imperfect, 
so the strengths of weak emission lines may not be reliable.
I present reduced spectra for 200 sources, which are available in the online 
supplementary material. I have excluded data that have limited wavelength
coverage or inadequate signal-to-noise ratios (SNRs) or background subtraction
for spectral classifications.

\begin{table}
\centering
\caption{Spectral and membership classifications of NIRSpec targets in the
ONC. The format and content are described here. The full table is
available in the online supplementary material.}
\label{tab:spec}
\begin{threeparttable}
\begin{tabular}{ll}
\hline
Column Label & Description\\
\hline
ID & Source number in catalog for JWST program 2770\\
Name & Other source name \\
RAdeg & Right ascension (ICRS)\tnote{\textit{a}}\\
DEdeg & Declination (ICRS)\tnote{\textit{a}}\\
SpType & Spectral type\\ 
r\_SpType & Spectral type reference\tnote{\textit{b}} \\
ak & $A_K$ from NIRSpec\\
member & ONC member?\\
evidence & Evidence of Membership\tnote{\textit{c}}\\
\hline
\end{tabular}
\begin{tablenotes}
\item[\textit{a}] Coordinates are from the reduced NIRCam images in
\citet{luh24o2}.
\item[\textit{b}] (1) this work, (2) \citet{wei09}, (3) \citet{hil13},
(4) \citet{luc06}, (5) \citet{rid07}, (6) \citet{hil97}, (7) \citet{ing14},
(8) \citet{sle04}, (9) \citet{luh24o1}.
\item[\textit{c}] All sources classified as nonmembers lack late-type
spectral features and evidence of youth and some are resolved galaxies
in the NIRCam images.
\end{tablenotes}
\end{threeparttable}
\end{table}

\section{Spectral Classifications}
\label{sec:class}

I have attempted to classify each NIRSpec target as an ONC member or
background source based on its spectrum and other available data.
To do that, I have used spectral diagnostics of age for late-type objects,
most notably the shape of the $H$-band continuum and the
strength of the CO band at 4.4--5.2~\micron. Young late-type objects have
triangular $H$-band continua and weak CO bands while older field dwarfs
have flat-topped $H$-band continua and strong CO absorption 
\citep{luc01,luh23}. In addition, I have employed signatures of youth in 
the form of IR excess emission from disks, emission lines from accretion
and ionization fronts surrounding disks, known as ``proplyds", and X-ray
emission \citep{get05,get22}.  Some targets are also 
resolved as proplyds, silhouette disks, or galaxies in the NIRCam images. 
Targets that lack M/L-type
spectral features and evidence of youth are classified as background sources. 
The 200 targets with useful spectra are classified as 53 members,
134 background sources, and 13 objects with uncertain membership.
Among the members, four are previously known proplyds \citep{ric08} and 
three may be new proplyds based on their emission lines and extended emission.
Two of the sources that are counted as members are dominated by emission 
lines and CO emission, indicating that they may be Herbig-Haro (HH) objects 
rather than stars.

For objects that are classified as ONC members, I have measured 
spectral types and extinctions through comparison of the NIRSpec data at
$<2.5$~\micron\ to standard spectra for young stars and brown dwarfs 
\citep[M0--L7,][]{luh17}. 
Several sources have uncertain spectral types because of low SNRs,
poor background subtraction, strong veiling, or very high reddenings.
A few stars are classified as members based on spectra from previous studies
but lack classifications from this work because of large gaps in the
NIRSpec wavelength coverage due to the edges of the detectors. 
Two spectra exhibit tentative detections of the
3.4~\micron\ absorption feature that has been observed in the faintest
substellar members of IC 348 \citep{luh24ic,luh25}, which has been attributed
to an unidentified aliphatic hydrocarbon. \citet{luh25} proposed the
assignment of a new spectral class of ``H" for such sources.
Aside from those two objects, the spectral classifications for ONC members
range from early M to early L. Twenty-four members have spectral types that 
are suggestive of brown dwarfs for the age of the ONC \citep[$>$M6,][]{bar15}.

The spectral and membership classifications for the 200 NIRSpec targets are 
presented in Table~\ref{tab:spec}. I have included spectral types from 
previous studies, which are available for 25 sources (23 members), and 
extinction estimates derived from the NIRSpec data for the ONC members that 
have well-defined spectral types. The source numbers in Table~\ref{tab:spec} 
are from the catalog utilized by program 2770 in the Astronomer's Proposal 
Tool (APT).

\begin{figure}
\includegraphics[width=\columnwidth]{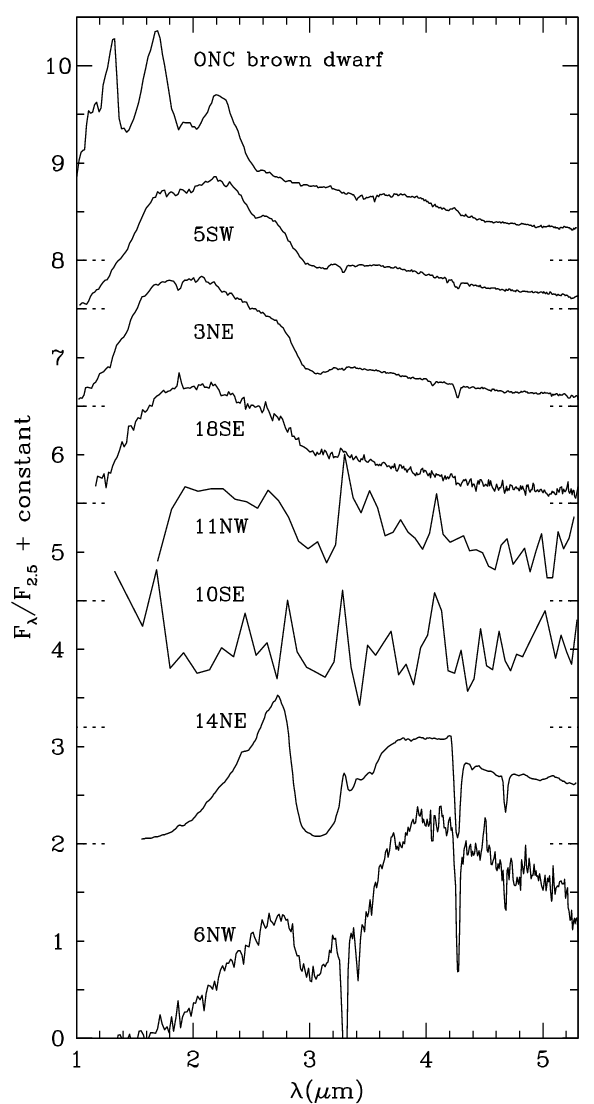}
\caption{JWST/NIRSpec spectra of an example of an ONC brown dwarf 
(source 784 in Table~\ref{tab:spec}) and seven components of JuMBOs from
\citet{pea23}, which are labeled with the source numbers from that study.
None of the JuMBOs exhibit the molecular absorption bands expected for 
brown dwarfs. For each spectrum that has been shifted upward, the
level for zero flux is marked by a pair of dotted tick marks.}
\label{fig:spec1} 
\end{figure}

\begin{figure*}
\includegraphics[width=\textwidth]{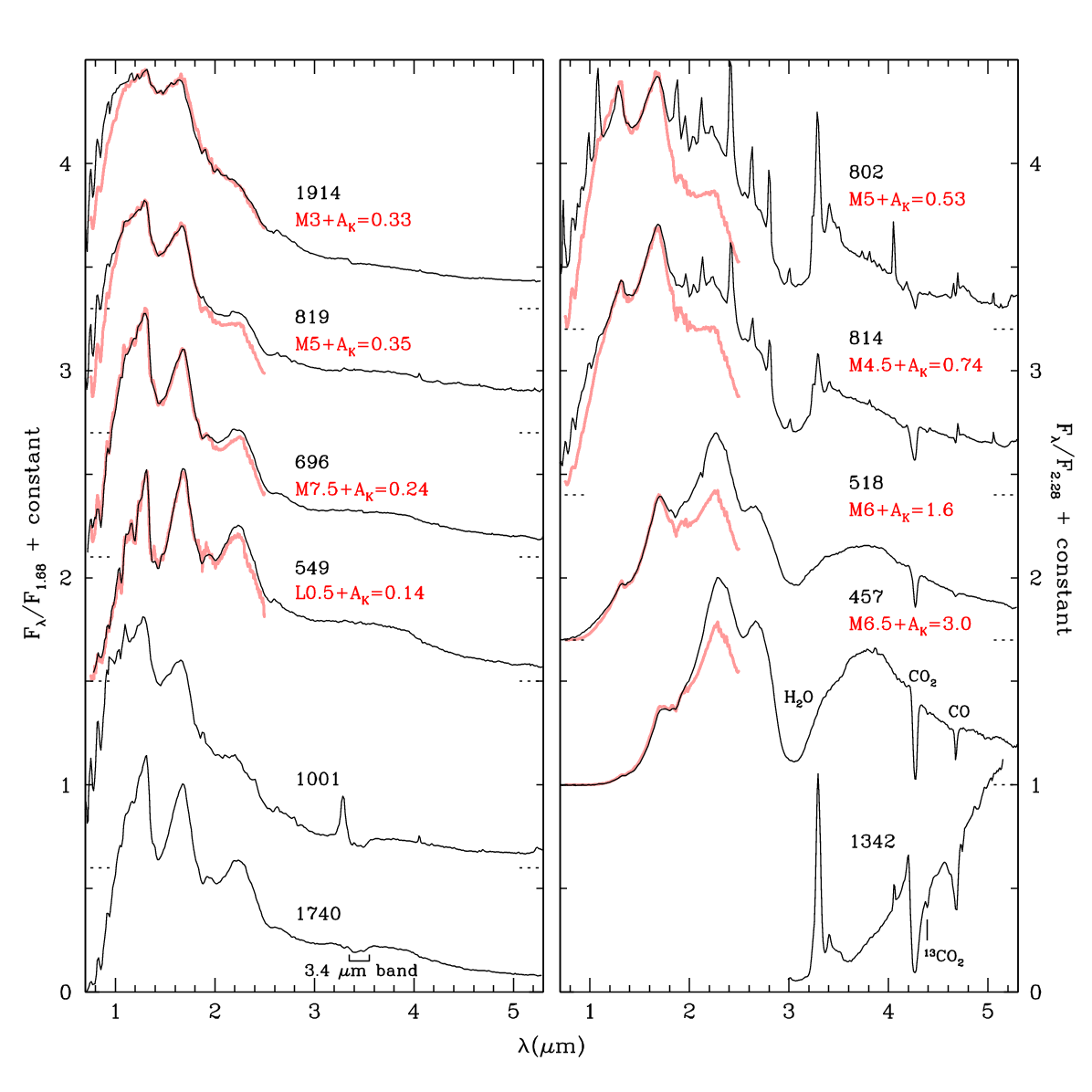}
\caption{Examples of JWST/NIRSpec spectra of sources classified as members of
the ONC. They are labeled with the source numbers from the APT catalog for 
these observations (Table~\ref{tab:spec}).  Some of the spectra are compared
to young standard spectra that have been reddened to match the slopes from 
1.2--1.7~\micron\ \citep{luh17}. Standards are not shown for the two objects 
that exhibit the 3.4~\micron\ feature (1001 and 1740) and the reddest
source (1342). For each spectrum that has been shifted upward, the level 
for zero flux is marked by a pair of dotted tick marks. The spectrum of 
source 1342 is normalized at 5~\micron.}
\label{fig:spec2} 
\end{figure*}

Seven of the NIRSpec targets consist of a single component of each
of seven JuMBO pairs from PM23. Their spectra are shown in 
Figure~\ref{fig:spec1} with an example of a NIRSpec target classified as 
an ONC brown dwarf. The data for sources 10SE and 11NW from PM23 have been 
binned to a lower resolution to increase their SNRs. The seven JuMBO components 
are classified as background sources based on the absence of the strong 
molecular absorption bands at $<$2.5~\micron\ (e.g, H$_2$O) that are 
observed in young brown dwarfs like the one included in Figure~\ref{fig:spec1}
(see also Figure~\ref{fig:spec2}). Most of the JuMBO spectra show a broad 
absorption feature near 3~\micron, which arises from H$_2$O ice in the 
molecular cloud and tends to be stronger at higher reddenings (for both 
cluster members and background stars). One of the JuMBOs, source 6NW, has 
large negative residuals at 3.3 and 3.4~\micron, which reflect imperfect 
subtraction of background PAH emission features.

In Figure~\ref{fig:spec2}, I present examples of spectra of sources that 
are classified as members of the ONC, consisting of four
that have low extinctions (mid-M to early L), the two that exhibit the 
3.4~\micron\ feature,
two that have strong emission lines (known and candidate proplyds),
and three that are highly reddened. Some of the spectra are compared to 
standards at the best matching spectral type and reddening.  The standards 
are reddened to match the ONC objects at 1.2--1.7~\micron\ according 
to the extinction law from \citet{sch16av}. Source 1342 is the reddest 
object in the NIRSpec sample and is sufficiently red that it could be a 
class 0 protostar \citep{fed24}. Among the 12 bands of NIRCam images from
MP23, it is not detected in F335M and at shorter wavelengths and is saturated 
in F444W. F360M and F470N are the only filters that have nonsaturated 
detections. The background near source 1342 in F444W is very bright,
so saturation occurs more easily (i.e., at a fainter magnitude) than in
most other areas of the image.
The blue and red wings of the CO$_2$ band in source 1342 
are elevated and suppressed, respectively, which is expected for large grain 
sizes \citep{dar22}. In the spectral image, strong PAH emission (3.3~\micron) 
is localized on the object, and the same appears to be true for the He~I
triplet at 1.083~\micron. PAH emission has been weak or absent in previous
observations of protostars \citep{van25}.

\begin{figure*}
\includegraphics[width=\textwidth]{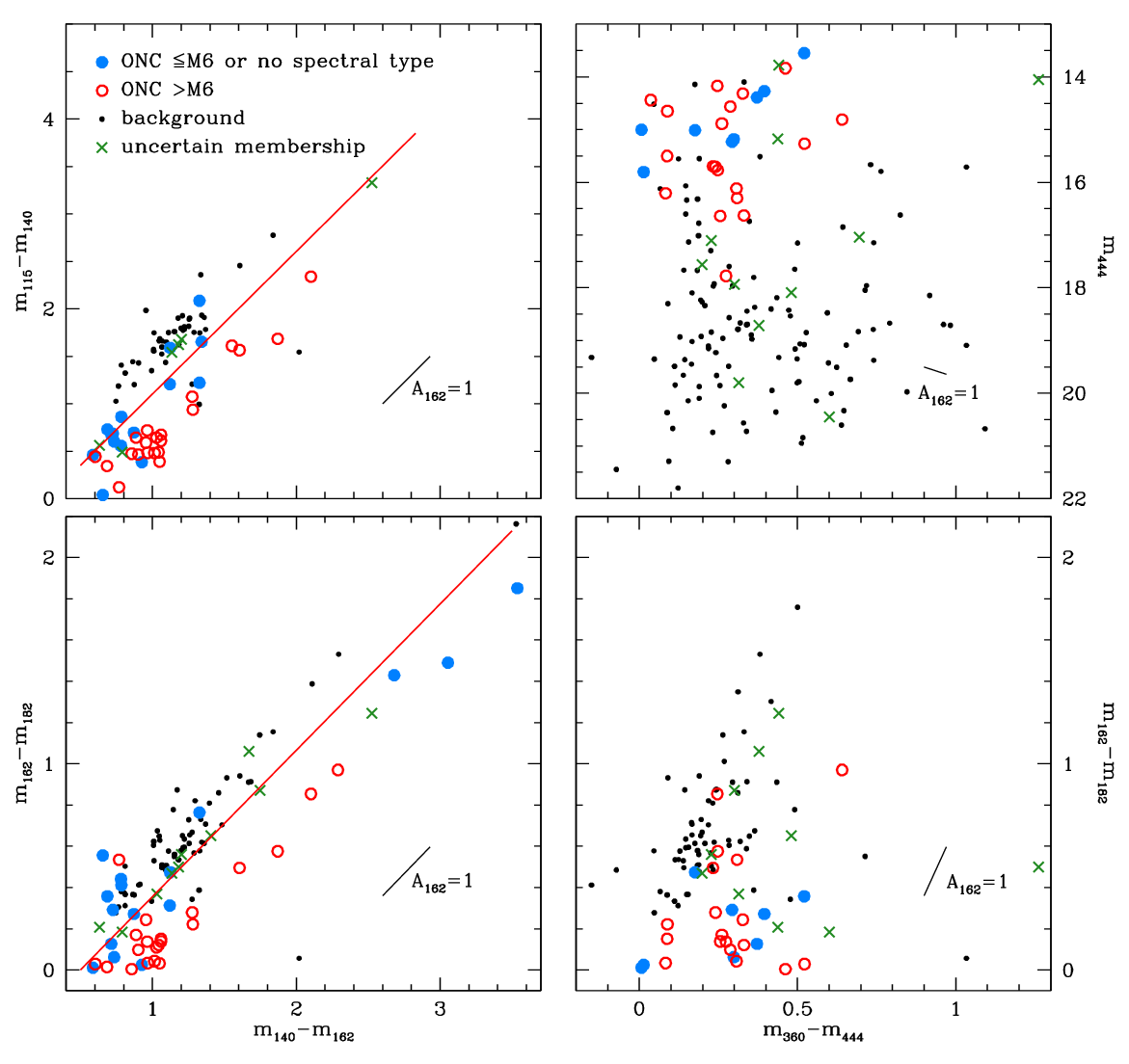}
\caption{Color-color and color-magnitude diagrams for JWST/NIRSpec targets
toward the ONC, which have been classified as members of the ONC (large filled
and open circles), background sources (small points), or objects with
uncertain membership (crosses; Table~\ref{tab:spec}). The photometry was
measured from JWST/NIRCam images \citep{luh24o2}.
For reference, I have marked the thresholds used in that study for
identifying brown dwarf candidates (red lines).}
\label{fig:cmd} 
\end{figure*}

\section{Testing Photometric Identification of Brown Dwarf Candidates}

The spectral classifications of the NIRSpec data are useful for testing
the recent selection of brown dwarf candidates in the ONC using the
NIRCam images \citep[PM23,][]{luh24o2}.

The JuMBOs from PM23 consist of 42 pairs and trios that have 86 components. 
A few of the components have spectroscopy from previous work.
Source 24 from PM23 is a $0\farcs1$ pair in which the components have similar 
magnitudes. Unresolved spectroscopy for the pair has produced a spectral 
type of M5.5 \citep{sle04}.
The brighter component of source 29 from PM23 was classified as M8 by
\citet{luh24o1}. Evolutionary models suggest that spectral types of M5.5
and M8 correspond to masses of $\sim0.1$ and 0.04~$M_\odot$ at the age of the
ONC \citep{bar15,cha23}, which are higher than the mass estimates of 
0.011--0.012~$M_\odot$ (11--12~$M_{\rm Jup}$) from PM23.  
As discussed in Section~\ref{sec:class}, 
seven JuMBO components are in the new NIRSpec sample, all of which are 
classified as background sources rather than brown dwarfs
(Figure~\ref{fig:spec1}).

The NIRSpec sample contains 24 brown dwarf candidates ($>$M6) identified in
\citet{luh24o2}. According to the NIRSpec data, they include 
five mid-M members (four are highly reddened),
12 $\geq$M6 members, one possible HH object, and two background sources. 
The remaining four candidates have uncertain membership; two have late types 
but low signal SNRs and two have mid-M types and very high reddenings.
I have plotted the NIRCam photometry from \citet{luh24o2} for the NIRSpec
sample in color-color and color-magnitude diagrams in Figure~\ref{fig:cmd}. 
Some sources are absent from a given diagram because of nondetection or
saturation in one of the relevant bands.
In the two diagrams on the left in Figure~\ref{fig:cmd}, I have included 
the boundaries (reddening vectors) that were used in \citet{luh24o2} for 
selecting brown dwarf candidates. All but one of the $>$M6 members in
the NIRSpec sample fall below those boundaries. The exception is source 3108,
whose unusually red value of $m_{162}-m_{182}$ is likely caused by its
strong Pa-$\alpha$ emission. Some $<$M6 members appear below the boundaries
as well; one is a likely HH object and the others have types of M5--M6
or are highly reddened.
Most objects that are classified as background sources with NIRSpec
are located above the boundaries, and thus were not expected to be brown
dwarfs based on the photometric analysis in \citet{luh24o2}. 
The color-magnitude diagram in Figure~\ref{fig:cmd} shows that the ONC
members are concentrated at brighter magnitudes in the NIRSpec sample.
At fainter magnitudes, only two late-type objects are present, both of which
have uncertain membership and spectral types because of low SNRs.

\section{Conclusions}

I have analyzed low-resolution 1--5~\micron\ spectra of 200 sources toward
the ONC that were obtained with JWST/NIRSpec through program 2770 in early
2024 (PI: M. McCaughrean). These data have been used to assess the membership 
of the sources in the ONC and to measure their spectral types.
I have classified 53 NIRSpec targets as likely cluster members, 24 of which
have spectral types that are suggestive of brown dwarfs ($>$M6)
and 23 of which have spectral classifications from previous work.
One new member could be a class 0 protostar and three others are possible
proplyds. Seven of the NIRSpec targets are candidates for Jupiter-mass objects
in binary systems identified by PM23 (JuMBOs), all of which
are classified as background sources with the NIRSpec data.
These spectral classifications support the results of my recent study of 
the JWST photometry in the ONC \citep{luh24o2}, which found that only 
a few JuMBO components are likely to be substellar, and that
none of those viable candidates form pairs that have uniquely 
wide separations or low masses relative to known binary brown dwarfs
\citep{cha04,bur07,tod14}.

\section*{Acknowledgements}

I thank Konstantin Getman, Tom Megeath, and Ewine van Dishoeck for helpful 
discussions. This work is based on observations made with the NASA/ESA/CSA 
James Webb Space Telescope. The data are associated with program 2770
and were obtained from MAST at the Space Telescope Science Institute, 
which is operated by the Association of Universities for Research 
in Astronomy, Inc., under NASA contract NAS 5-03127. 
The Center for Exoplanets and Habitable Worlds is supported by the
Pennsylvania State University, the Eberly College of Science, and the
Pennsylvania Space Grant Consortium.

\section*{Data Availability}

The data in Table~\ref{tab:spec} and the JWST spectra for those sources
are available in the online supplementary material. 

\bibliographystyle{mnras}
\bibliography{ref} 

\bsp	
\label{lastpage}
\end{document}